# REQUIREMENTS OF VERTICAL HANDOFF MECHANISM IN 4G WIRELESS NETWORKS

Mandeep Kaur Gondara[1] and Dr. Sanjay Kadam[2]

[1] Ph. D Student, Computer Science Department, University of Pune, Pune
`u08401@cs.unipune.ac.in`

[2] Research Guide, Computer Science Department, University of Pune, Pune
`sskadam@cdac.in`

## ABSTRACT

*The importance of wireless communication is increasing day by day throughout the world due to cellular and broadband technologies. Everyone around the world would like to be connected seamlessly anytime anywhere through the best network. The 4G wireless system must have the capability to provide high data transfer rates, quality of services and seamless mobility. In 4G, there are a large variety of heterogeneous networks. The users for variety of applications would like to utilize heterogeneous networks on the basis of their preferences such as real time, high availability and high bandwidth. When connections have to switch between heterogeneous networks for performance and high availability reasons, seamless vertical handoff is necessary. The requirements like capability of the network, handoff latency, network cost, network conditions, power consumption and user's preferences must be taken into consideration during vertical handoff. In this paper, we have extracted the requirements of a vertical handoff from the literature surveyed. The evaluation of the existing work is also being done on the basis of required parameters for vertical handoff. A sophisticated, adaptive and intelligent approach is required to implement the vertical handoff mechanism in 4G wireless networks to produce an effective service for the user by considering dynamic and non dynamic parameters.*

## 1. INTRODUCTION

Mobility is the most important feature of today's wireless networking system. Mobility can be attained by handoff mechanisms in wireless networks. Handoff is the process of changing the channel (frequency, time slot, spreading code, or combination of them) associated with the current connection while a call is in progress [1].

### 1.1 Types of Handoffs in 4G Networks

In 4G networks, the handoffs are classified into two main streams

#### 1.1.1 Horizontal Handoff

Handoff between two base stations (BSs) of the same system is called Horizontal handoff. Horizontal handoff involves a terminal device to change cells within the same type of network (e.g., within a CDMA network) to maintain service continuity [2]. It can be further classified into Link-layer handoff and Intra-system handoff. Horizontal handoff between two BS, under same foreign agent (FA) is known as Link-layer handoff. In Intra-system handoff, the

International Journal of Wireless & Mobile Networks (IJWMN) Vol. 3, No. 2, April 2011

horizontal handoff occurs between two BSs that belong to two different FAs and both FAs belongs to the same system and hence to same gateway foreign agent (GFA).

### 1.1.2  Vertical Handoff (VHO)

Vertical handoff refers to a network node changing the type of connectivity it uses to access a supporting infrastructure, usually to support node mobility. For example, a suitably equipped laptop might be able to use both a high speed wireless LAN and a cellular technology for Internet access. Wireless LAN connections generally provide higher speeds, while cellular technologies generally provide more ubiquitous coverage. Thus the laptop user might want to use a wireless LAN connection whenever one is available, and to 'fall over' to a cellular connection when the wireless LAN is unavailable. Vertical handovers refer to the automatic fallover from one technology to another in order to maintain communication [3].The vertical handoff mechanism allows a terminal device to change networks between different types of networks (e.g., between 3G and 4G networks) in a way that is completely transparent to end user applications[2].

The vertical handoff process involves three main phases [4], [5], namely system discovery, vertical handoff decision, and VHO execution. During the system discovery phase, the mobile terminal determines which networks can be used. These networks may also advertise the supported data rates and Quality of Service (QoS) parameters. In VHO decision phase, the mobile terminal determines whether the connections should continue using the current network or be switched to another network. The decision may depend on various parameters or metrics including the type of the application (e.g., conversational, streaming), minimum bandwidth and delay required by the application, access cost; transmit power, and the user's preferences. During the VHO execution phase, the connections in the mobile terminal are re-routed from the existing network to the new network in a seamless manner. This phase also includes the authentication, authorization, and transfer of a user's context information [6].
Handoff management aims at controlling the change of an access point (AP) in order to maintain the connection with the moving device during the active data transmission. The problem is exacerbated by the presence of APs adopting different technologies. Hence vertical handoffs, that is, handoff procedures between APs of heterogeneous technology, should be taken into account [7].

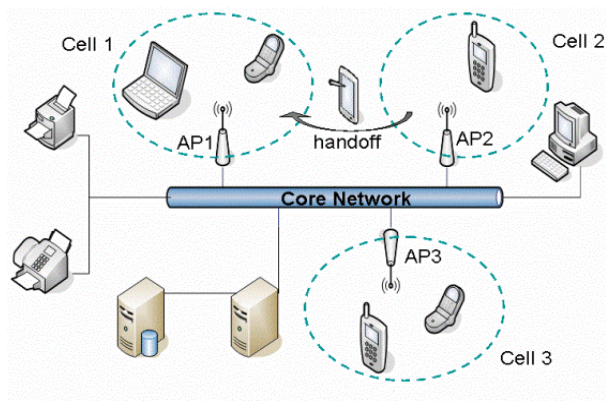

Figure 1: Vertical Handoff in heterogeneous networks

In this paper, we propose requirements for vertical handoff decision model for heterogeneous 4G networks on the basis of literature surveyed. Section 4 represents dynamic and non dynamic parameters for VHO (vertical handover) mechanism in 4G networks. In section 5, the evaluation of existing work is being done on the basis of parameters discussed in section4.

## 2. STATE OF THE ART

From the literature surveyed, different authors use different terms such as models, techniques and approaches to refer mechanisms. In order to compile a VHO mechanism for 4G wireless networks, it is essential to study existing VHO mechanisms. The study of existing mechanisms will assist in the identification of requirements for VHO mechanism. As of now, a few approaches for VHO have been found in the literature. One kind of approach is based on "Received Signal Strength (RSS)" that may be combined with other parameters such as network load and network cost. The another kind of approaches are using artificial intelligence techniques, combining several parameters such as network conditions and Mobile Terminal's (MT) mobility in the handoff decision[9]. Some are policy based approaches, combining several metrics such as access cost, power consumption, and bandwidth, velocity of a host, quality of service in VHO mechanism.

. Wang et al. [8] have introduced the policy enabled handoff (PEHHWN). In their work, they describe a policy-enabled handoff system that allows users to express policies and to find out the best network on the basis of dynamic and static parameters such as network cost, performance and power consumption. However, the cost function presented in that paper is very preliminary and cannot handle sophisticated configurations. Another policy based work is proposed in [11], where the Automatic Handover Manager (AHM) provides a solution for determining the best network interfaces for the services (TAHDM). The decision is made by using the context information from the mobile node, networks and the user as well as the RSS. AHM is based on the autonomic computing concept. It provides a good policy for the vertical handover using the context information without user's interventions. AHM has four major functions such as monitoring, analyzing, planning and executing. This paper describes how to compose the context evaluation function and formulate a policy. In future work, more concrete context information and improvements can be made in AHM by optimizing the context evaluation function. In [16], the objective of research work is to determine the conditions under which vertical handoff should be performed for heterogeneous wireless networks (VHDAHWN). This work incorporated the connection duration and signalling load incurred on the network for VHO decision. The work is based on the Markov decision process (MDP) formulation to maximize the expected total reward of a connection. Numerical results show that their proposed MDP algorithm gives a higher expected total reward and lower expected number of vertical handoffs than SAW (Simple Additive Weighting) and GRA (Grey Relational Analysis), and two heuristic policies under a wide range of conditions. Their proposed model is adaptive and applicable to a wide range of conditions.

A. Dvir et al. [13] proposed an efficient decision handoff mechanism for heterogeneous network (EDHMHN). A decision function in which the system considers all the available network and user parameters (e.g. host velocity, battery status, Wi-Fi AP's current load, and WiMAX BS's Qos guarantees, and performs technology selection such that an overall system performance metric is optimized(i.e., throughput and capacity limitation). They have defined a new system-wise entity that is activated when a user is in an area with overlapping access technologies and needs to decide the best technology to be used, where the entity performs technology selection in order to optimize the overall system performance metric in terms of throughput and capacity limitation. Their simulation results validate the efficiency of their method and show that it is also applicable to other combinations of access technologies.

Mrs. Chandralekha et al. [12] proposed a theory for selection of the best available wireless network during handoffs based on a set of predefined user preferences on a mobile device (UARTVHO). A neural network model has been used to process multi-criteria VHO decision metrics. The features used from generated data have been carefully selected and used as inputs for the neural network in order to have high performance rate. A modified type of competitive learning called "Adaptive Resonance Theory (ART)" has been designed to overcome the problem of learning stability. The proposed method is capable of selecting the best available wireless network with a reasonable performance rate. The overall approach is based on artificial intelligence, combining some other metrics for decision model of VHO.

Goyal et al. [9] proposed a dynamic decision model for VHO across heterogeneous wireless networks (ADDMVHO). This model makes the right VHO decisions by determining the "best" network at "best" time among available networks based on dynamic factors such as RSS and velocity of mobile station as well as static factors. A handoff Management Center (HMC) monitors the various inputs collected from the network interfaces and their base stations (BS) analyze this information and make handoff decisions. The dynamic algorithm has different phases. The Priority Phase is used to remove all the unwanted and ineligible networks from the prospective candidate networks. The Normal Phase is used to accommodate user-specific preferences regarding the usage of network interfaces. Finally, the Decision Phase is used to select the "Best" network and executing the handoff to the selected network. In [10], the proposed research provides optimized performance in heterogeneous wireless networks during VHO decision (VHDAPOP). A VHO decision algorithm is being developed that enables a wireless access network to balance the overall load among all attachment points (e.g., Base Stations (BSs) and Access Points (APs) and also to maximize the collective battery lifetime of mobile nodes (MNs). In addition, when ad hoc mode is applied to 3/4G wireless data networks, a route selection algorithm has been devised to forward data packets to the most appropriate attachment point for maximizing the collective battery lifetime and to maintain load balancing.

## 3. SOURCES OF SURVEY

The requirements of a VHO mechanism for wireless networks extracted from the literature surveyed. In the research, we used documentary sources, which involve existing textual documents available in electronic and printed media. The data sources used in this research include academic journals, applicable books and the internet.

In this survey, textual analysis is used as means of data collection. Textual analysis involves both content analysis and textual interpretations. Based on the research departure points, namely the vertical handoff mechanism and parameters of handoffs in wireless networks, the contents of the referenced publications were analyzed to find their applicability to the study. The requirements were extracted from existing handoff decision models and mechanisms and other surveyed literature on handoff aspects in the wireless network environment.Different authors have indicated different aspects that should be considered while designing a handoff mechanism for the wireless networks. These aspects include many parameters as discussed in section 4 for seamless and secure handoffs in wireless environment.

## 4. REQUIREMENT FOR HANDOFF MECHANISM

### 4.1 Bandwidth

Bandwidth is a measure of the width of a range of frequencies. It is the difference between the upper and lower frequencies in a contiguous set of frequencies. In order to provide seamless handoff for Quality of service (Qos) in wireless environment, there is a need to manage bandwidth requirement of mobile node during movement. Bandwidth is generally known as the link capacity in a network. Higher offered bandwidth ensures lower call dropping and call blocking probabilities; hence higher throughput [9].Bandwidth handling should be an integral part of any of the handoff technique.

### 4.2 Handoff Latency

Handover of calls between two BS is encountered frequently and the delay can occur during the process of handoffs. This delay is known as handoff latency. A good handoff decision model should consider Handoff latency factor and the handoff latency should be minimized. Many proposed handoff decision models have tried to minimize the handoff latency by incorporating this factor in their handoff decision models. Handoff Latencies affect the service quality of many applications of mobile users. It is essential to consider handoff latency while designing any handoff technique.

### 4.3 Power Consumption

In 4G networks, we need to find ways to improve energy efficiency. Power is not only consumed by user terminal but also attributed to base station equipments. Power is also consumed during mobile switching or handoffs. During handoff, frequent interface activation can cause considerable battery drainage. The issue of power saving also arises in network discovery because unnecessary interface activation can increase power consumption. It is also important to incorporate power consumption factor during handoff decision.

### 4.4 Network Cost

A multi criteria algorithm for handoff should also consider the network cost factor. The cost is to be minimized during VHO in wireless networks. The new call arrival rates and handoff call arrival rates can be analyzed using cost function. Next Generation heterogeneous networks can combine their respective advantages on coverage and data rates, offering a high Quality of Service (QoS) to mobile users. In such environment, multi-interface terminals should seamlessly switch from one network to another in order to obtain improved performance or at least to maintain a continuous wireless connection. Therefore, network selection cost is important in handoff decisions.

### 4.5 User Preferences

When handover happens, the users have more options for heterogeneous networks according to their preferences and network performance parameters. The user preferences could be preferred networks, user application requirements (real time, non-real time), service types (Voice, data, video), Quality of service (It is a set of technologies for managing network traffic in a cost effective manner to enhance user experiences for wireless environments) etc. User Preferences can also be considered for VHO in 4G wireless networks.

## 4.6 Network Throughput

Network throughput refers to the average data rate of successful data or message delivery over a specific communications link. Network throughput is measured in bits per second (bps). Maximum network throughput equals the TCP window size divided by the round-trip time of communications data packets. As network throughput is considered in dynamic metrics for making decision of VHO, it is one the important requirement to be considered for the VHO.

## 4.7 Network Load Balancing

Network load is to be considered during effective handoff. It is important to balance the network load to avoid deterioration in quality of services. Variations in the traffic loads among cells will reduce the traffic-carrying capacity. To provide a high quality communication service for mobile subscribers and to enhance a high traffic-carrying capacity when there are variations in traffic, network load must be paid attention.

## 4.8 Network Security

With the increasing demand of wireless networks, seamless and secure handoff has become an important factor in wireless networks. The network security consists of the provisions and policies adopted by the network to prevent and monitor unauthorized access, misuse, modification, and network-accessible resources. In a wireless environment, data is broadcast through the air and people do not have physical controls over the boundaries of transmissions. The security features provided in some wireless products may be weaker; to attain the highest levels of integrity, authentication, and confidentiality, network security features should be embedded in the handoff policies.

## 4.9 Received Signal strength (RSS)

The performance of a wireless network connection depends in part on signal strength. Between a mobile node (MN) and access point (AP), the wireless signal strength in each direction determines the total amount of network bandwidth available along that connection. RSS depicts the power present in a received signal. A signal must be strong enough between base station and mobile unit to maintain signal quality at receiver. The RSS should not be below a certain threshold in a network during handoff. VHO includes three sequential steps as discussed earlier in this paper, namely handoff initiation, handoff decision and handoff execution. Handoff initiation is concerned with measurement of RSS [14].

## 4.10    Velocity

Velocity of the host should also be considered during handoff decision. Because of the overlaid architecture of heterogeneous networks, handing off to an embedded network, having small cell area, when travelling at high speeds is discouraged since a handoff back to the original network would occur very shortly afterwards [9].

However, we have stated the important parameters/metrics as requirements but other parameters such as network conditions, network capability and bit error rate can also be considered during vertical handoff.  The dynamic requirements include RSS, velocity, throughput, user preferences as parameters and non-dynamic requirements include network cost, power consumption, network security and bandwidth as parameters. A good handoff mechanism decision model should have both dynamic and non-dynamic metrics. However, it is important to consider maximum number of static and dynamic requirements during VHO but it is difficult to include all the metrics in a single decision model due to complexity of algorithms and conflicting issues of multiple metrics.

## 5. EVALUATION OF THE EXISTING WORK

In this section, existing handoff mechanisms are evaluated against the requirements of a VHO in a wireless environment. The evaluation is intended to establish the gap between the existing handoff mechanisms and the handoff requirements for the wireless environment. In table 1 we are listing the handoff models for reference purpose in table 2. In table2, the evaluation of existing work is done against the requirements for VHO discussed in section4.

| Abbreviation | Ref# | Handoff Mechanisms |
|---|---|---|
| UARTVHO | 12 | Use of Adaptive Resonance Theory for Vertical Handoff Decision in heterogeneous Wireless Environment |
| VHDAHWN | 16 | A Vertical Handoff Decision Algorithm for Heterogeneous Wireless Networks |
| TAHDM | 11 | Towards Autonomic Handover Decision Management in 4G Networks |
| EDHMHN | 13 | Efficient decision handoff mechanism for heterogeneous network |
| PEHHWN | 8 | Policy-Enabled Handoffs across Heterogeneous Wireless Networks |
| VHDAPOP | 10 | Vertical Handoff Decision Algorithms for Providing Optimized Performance in heterogeneous Wireless Networks |
| ADDMVHO | 9 | A Dynamic Decision Model for Vertical Handoffs across Heterogeneous Wireless Networks |

Table 1: Abbreviations for the existing handoff mechanisms

| Handoff Mechanism | Requirement Number | | | | | | | | | |
|---|---|---|---|---|---|---|---|---|---|---|
| | 1 | 2 | 3 | 4 | 5 | 6 | 7 | 8 | 9 | 10 |
| UARTVHO | X | X | | X | X | X | X | X | | |
| VHDAHWN | X | | | X | X | X | X | | X | |
| TAHDM | X | | X | X | X | X | | | X | X |
| EDHMHN | X | | X | | X | X | X | | X | X |
| PEHHWN | X | X | X | X | | | X | | | |
| VHDAPOP | X | | X | X | X | | X | | | |
| ADDMVHO | X | | X | X | | | | | X | X |

Table 2: Requirements satisfied by the existing handoff mechanism

Using the requirement parameters as stated in section 4, it has been found that there is no vertical handoff mechanism that satisfies all the requirements, although all VHO models/mechanisms satisfy at least five requirements.

- For requirement 1, vertical handoff mechanisms discussed above indicate the importance of bandwidth in wireless networks. However, Vertical handoff models such as UARTVHO, VHDAHWN, EDHMHN, PEHHWN, VHDAPOP, and ADDMVHO have incorporated this requirement in their model but still a mechanism is required for controlling the variations in bandwidth while a MN is switching from high to low or low to high bandwidth network.

- For Requirement 2, the delay during the handover process is to be minimized. Handoff mechanisms such as UARTVHO and VHDAHWN have tried to minimize the handoff latency by incorporating this factor in their handoff decision models.
- For requirement 3, there is a need to find ways to improve energy efficiency. The handoff models that have considered this requirement are TAHDM, EDHMHN, VHDAPOP, PEHHWN and ADDMVHO.
- For requirement 4, the network selection cost is to be incorporated in the decision model of a VHO. These includes UARTVHO, VHDAHWN, TAHDM, VHDAPOP, PEHHWN and ADDMVHO mechanisms which are using cost functions to analyze the network cost during switching of networks by a mobile node.
- For requirement 5, the user preferences must be considered in terms of preferred network, service types and requirements of applications. The frameworks that consider user preferences during handoff decision are UARTVHO, VHDAHWN, TAHDM, VHDAPOP and PEHHWN.
- Requirement 6 states that the VHO mechanism should consider average data rate of successful data over a communication link. Handoff mechanisms such as UARTVHO, VHDAHWN, TAHDM and EDHMHN are able to satisfy this requirement.
- For requirements 7, it is important to balance the network load for traffic carrying capacity, quality of services and for providing high quality communication. Many of the proposed handoff mechanisms satisfy this requirement such as VHDAHWN, VHDAPOP, UARTVHO, EDHMHN, and PEHHWN.
- For requirement 8, secure handoff has become an important factor in wireless networks. Network security feature must be merged along with other parameters in the decision model of VHO. As in table 2, only one work "UARTVHO" considers this parameter. There is a need to integrate network security during handoff.
- Requirement 9 states that the signal strength plays crucial role in performance of the wireless network by depicting the power present in a signal. The state of the art include VHDAHWN,TAHDM, EDHMHN and ADDMVHO work where handoff decision model is based on the RSS.
- For requirement 10, velocity of the host must be paid attention during handoff decision. The handoff mechanisms such as EDHMHN, PEHHWN and ADDMVHO consider this requirement.

However, the evaluation of the existing work is being done on the basis of requirements extracted from the literature surveyed but the evaluation is not being done on the basis of performance of the algorithms. There are basically three mechanisms used to evaluate the performance of the handoff mechanisms, analytical, simulation and emulation approaches as presented in the literature. The analytical approach works well under certain specified constraints. The simulation approach is used commonly to verify the experimental results on the simulator. The emulation approach uses a software simulator consisting of handoff algorithms to process measured variables (for instance, received signal strength and bit error rate) [17]. Before deciding the parameters for decision mechanism of VHO, the performance based evaluation can be done for better results.

The decision mechanism of VHO can become more fruitful, if the number of parameters is more during decision making as stated in section 4. The success of vertical handoff mechanism depends upon the decision model based on requirements/metrics. An efficient VHO decision mechanism can not only enhance the system capacity but also improve the quality of services for a user. The existing works can be extended or new works can be developed to incorporate more parameters in VHO mechanism.

## 6. CONCLUSION

The vertical handoff will remain an essential component for 4G wireless networks due to switching of mobile users amongst heterogeneous networks. In this paper, we have described a few works in vertical handoff mechanisms and exposed a summary of decision algorithms for vertical handoff in literature. The 4G wireless networks create new handoff challenges due to multiple requirements for vertical handoff. In this paper, the requirements of a vertical handoff for 4G wireless network were proposed. The requirements include high bandwidth, low handoff latency, lower power consumption, minimum network cost, balanced network load, network security, user preferences, throughput and RSS of a switching network. Establishing the requirements of a vertical handoff mechanism for 4G wireless networks is a critical milestone in the development of vertical handoff mechanism for 4G. In this paper, the evaluation of existing vertical handoff mechanisms is also done against the requirements stated in the paper. The evaluation indicate the need to have a VHO mechanism for 4G wireless networks that has the ability to satisfy maximum number of requirements. However, it is difficult to consider all the parameters during designing the decision model for VHO but if we consider more parameters, the outcome of the decision mechanism would definitely improve.